\documentclass[acus]{JAC2001}

\usepackage{graphicx,amssymb}

\begin{document}

{
\onecolumn \thispagestyle{empty}
\renewcommand{\thefootnote}{\fnsymbol{footnote}}
\setcounter{footnote}{0}

%\end{document}

\begin{flushright}
{\normalsize
SLAC-PUB-9226\\
arXiv:physics/0206002\\
June 2002}
\end{flushright}

\vspace{.8cm}

\begin{center}
{\bf\Large A Simplified Model of Intrabeam Scattering
}\footnote{Work supported by Department of Energy contract
DE--AC03--76SF00515.}

\vspace{1cm}

{\large
K.L.F. Bane\\
{\it Stanford Linear Accelerator Center, Stanford University,}\\
{\it Stanford, CA 94309 USA}}

\end{center}

\vspace{.5cm}

\abstract{ Beginning with the general Bjorken-Mtingwa solution, we
derive a simplified model of intrabeam scattering (IBS), one valid
for high energy beams in normal storage rings; our result is
similar, though more accurate than a model due to Raubenheimer. In
addition, we show that a modified version of Piwinski's IBS
formulation (where $\eta^2_{x,y}/\beta_{x,y}$ has been replaced by
${\cal H}_{x,y}$) at high energies asymptotically approaches the
same result. }

\vfill

\begin{center}
{\it Presented at the}
{\it Eighth European Particle Accelerator Conference (EPAC'02),} \\
{\it Paris, France}\\
{\it June 3-7, 2002}
\end{center}

} \setcounter{footnote}{0}

%%
%%   VARIABLE HEIGHT FOR THE TITLE BOX (default 35mm)
%%

\setlength{\titleblockheight}{25mm}

\title{A SIMPLIFIED MODEL OF INTRABEAM SCATTERING}

\author{K.L.F. Bane, SLAC, Stanford, CA94309, USA
\thanks{Work supported by the Department
of Energy, contract DE-AC03-76SF00515}
}

\maketitle

\begin{abstract}

Beginning with the general Bjorken-Mtingwa solution,
we derive a simplified model
of intrabeam scattering (IBS), one valid
for high energy beams in normal storage
rings; our result is similar, though more accurate
than a model due to Raubenheimer. In addition, we
show that a modified version of Piwinski's
IBS formulation (where $\eta^2_{x,y}/\beta_{x,y}$ has been
replaced by ${\cal H}_{x,y}$) at high energies
asymptotically approaches the same
result.

\end{abstract}
\bibliographystyle{unsrt}

\section{INTRODUCTION}

Intrabeam scattering (IBS), an effect that tends to increase the
beam emittance, is important in hadronic\cite{Bhat:99} and heavy
ion\cite{RHIC:01} circular machines, as well as in low emittance
electron storage rings\cite{ATF:02B}. In the former type of
machines it results in emittances that continually increase with
time; in the latter type, in steady-state emittances that are
larger than those given by quantum excitation/synchrotron
radiation alone.

The theory of intrabeam scattering for accelerators was first developed by
Piwinski\cite{Piwinski:74}, a result that was extended by
Martini\cite{Martini:84}, to give a formulation that we call here
the standard Piwinski~(P) method\cite{Piwinski:99}; this was
followed by the equally detailed Bjorken and Mtingwa (B-M)
result\cite{Bjorken:83}. Both approaches solve the local,
two-particle Coulomb scattering problem for (six-dimensional)
Gaussian, uncoupled beams, but the two results appear to be
different; of the two, the B-M result is thought to be the more
general\cite{Piwinski:p}.

For both the P and the B-M methods solving
for the IBS growth rates is time consuming,
involving, at each time (or iteration) step, a numerical integration
at every lattice element.
Therefore, simpler, more approximate
formulations of IBS have been developed over the years: there are
approximate solutions of
Parzen\cite{Parzen:87}, Le Duff\cite{LeDuff:89},
Raubenheimer\cite{Raubenheimer:91}, and Wei\cite{Wei:93}.
In the present report we
derive---starting with the general B-M formalism---another
approximation, one valid for
high energy beams and more accurate than Raubenheimer's approximation.
We, in addition, demonstrate that under these same conditions
a modified version of Piwinski's IBS formulation asymptotically becomes equal to
this result.

%\section{HIGH ENERGY APPROXIMATION TO BJORKEN-MTINGWA}
\section{HIGH ENERGY APPROXIMATION}

\subsection{The General B-M Solution\cite{Bjorken:83}}

Let us consider bunched beams that are uncoupled, and include
vertical dispersion due to {\it e.g.} orbit errors. Let the
intrabeam scattering growth rates be
\begin{equation}
{1\over T_p}={1\over\sigma_p}{d\sigma_p\over dt}\ ,\quad {1\over
T_x}={1\over\epsilon_x^{1/2}}{d\epsilon_x^{1/2}\over dt}\ ,\quad
{1\over T_y}={1\over\epsilon_y^{1/2}}{d\epsilon_y^{1/2}\over dt}\ ,
\label{growth_def_eq}
\end{equation}
with $\sigma_p$ the relative energy spread, $\epsilon_x$ the
horizontal emittance, and $\epsilon_y$ the vertical emittance.
The growth rates according to Bjorken-Mtingwa
(including a $\sqrt{2}$ correction factor\cite{Kubo:01b}, and including
vertical dispersion) are
%\begin{widetext}
%\begin{equation}
%{1\over T_i}\ =\ 8\pi A({\rm log})\bigg<\int_0^\infty {d\lambda\,
%\lambda^{1/2}\over [{\rm det}(L+\lambda I)]^{1/2}}\
%\bigg\{ TrL^{(i)}Tr\left({1\over L+\lambda I}\right)
% -\ 3TrL^{(i)}
%\left({1\over L+\lambda I}\right)\bigg\}\bigg>\quad,
%\end{equation}
%\end{widetext}
\begin{eqnarray}
&&\hspace{-16pt}{1\over T_i}\ =\ 4\pi A({\rm log})\bigg<\int_0^\infty {d\lambda\,
\lambda^{1/2}\over [{\rm det}(L+\lambda I)]^{1/2}}\bigg\{
\nonumber\\
&&\hspace{-8pt}
 TrL^{(i)}Tr\left({1\over L+\lambda I}\right)
 -\ 3TrL^{(i)}
\left({1\over L+\lambda I}\right)\bigg\}\bigg>\quad\ \
\label{BM_eq}
\end{eqnarray}
where $i$ represents $p$, $x$, or $y$;
\begin{equation}
A= {r_0^2c N\over 64\pi^2\bar\beta^3\gamma^4\epsilon_x\epsilon_y\sigma_s\sigma_p}
\quad,
\end{equation}
with $r_0$ the classical particle radius,
$c$ the speed of light, $N$ the bunch population,
$\bar\beta$ the velocity over $c$, $\gamma$ the Lorentz energy factor,
and $\sigma_s$ the bunch length;
$({\rm log})$ represents the Coulomb log factor,
$\langle\rangle$ means that the enclosed
quantities, combinations of beam parameters and lattice
properties,
 are averaged around the entire ring; ${\rm det}$ and $Tr$ signify, respectively, the determinant and the
trace of a matrix, and $I$ is the unit matrix.
Auxiliary matrices are defined as
\begin{equation}
L=  L^{(p)} + L^{(x)} + L^{(y)} \quad,
\end{equation}
\begin{equation}
L^{(p)}={\gamma^2\over\sigma_p^2}\left(
\begin{array}{ccc}
0 & 0 & 0 \\
0 & 1 & 0 \\
0 & 0 & 0
\end{array}\right)\quad,
\end{equation}
\begin{equation}
L^{(x)}={\beta_x\over\epsilon_x}\left(
\begin{array}{ccc}
1 & -\gamma\phi_x & 0 \\
-\gamma\phi_x & {\gamma^2{\cal H}_x/\beta_x} & 0 \\
0 & 0 & 0
\end{array}\right)\quad,
\end{equation}
\begin{equation}
L^{(y)}={\beta_y\over\epsilon_y}\left(
\begin{array}{ccc}
0 & 0 & 0 \\
0 & {\gamma^2{\cal H}_y/\beta_y} & -\gamma\phi_y \\
0 & -\gamma\phi_y & 1
\end{array}\right)\quad.
\end{equation}
The dispersion invariant is
${\cal H}=[\eta^2+(\beta\eta^\prime-{1\over 2}\beta^\prime\eta)^2]/\beta$,
and $\phi=\eta^\prime-{1\over 2}\beta^\prime\eta/\beta$, where
$\beta$ and $\eta$ are the beta and dispersion lattice functions.

%For
%unbunched beams $\sigma_s$ in Eq.~\ref{BM_eq} is replaced
%by $C/(2\sqrt{2\pi})$, with $C$ the circumference of the machine.

\subsection*{The Bjorken-Mtingwa Solution at High Energies}

Let us first consider $1/T_p$ as given by Eq.~\ref{BM_eq}.
Note that if we change the integration variable
to $\lambda^\prime=\lambda\sigma_H^2/\gamma^2$ then
\begin{equation}
(L+\lambda^\prime I)={\gamma^2\over\sigma_H^2}\left(
\begin{array}{ccc}
a^2+\lambda^\prime & -a\zeta_x & 0 \\
-a\zeta_x & 1+\lambda^\prime & -b\zeta_y \\
0 & -b\zeta_y & b^2+\lambda^\prime
\end{array}\right)\quad,
\end{equation}
with
\begin{equation}
{1\over\sigma_H^2}= {1\over\sigma_p^2} +
{{\cal H}_x\over\epsilon_x} +
{{\cal H}_y\over\epsilon_y}\quad,
\end{equation}
\begin{equation}
a={\sigma_H\over\gamma}\sqrt{\beta_x\over\epsilon_x}\ ,\quad
b={\sigma_H\over\gamma}\sqrt{\beta_y\over\epsilon_y}\ ,\quad
\zeta_{x,y}=\phi_{x,y}\sigma_H\sqrt{\beta_{x,y}\over\epsilon_{x,y}}
\label{ab_rev2_eq}
\end{equation}
Note that, other than a multiplicative factor, there are only 4 parameters
in this matrix: $a$, $b$, $\zeta_x$, $\zeta_x$. Note that, since
$\beta\phi^2\leq{\cal H}$, the parameters $\zeta<1$; and that
if ${\cal H}\approx\eta^2/\beta$ then $\zeta$ is small. We give, in Table~1,
average values of $a$, $b$, $\zeta_x$, in selected electron rings.\vspace{-4mm}
%Note that for the NLC ring $\zeta_x$ periodically reaches up to the value 0.8.

\begin{table}[htb]
\begin{center}
\caption{
Average values of $a$, $b$, $\zeta_x$, in selected electron rings.
The zero current emittance ratio $\sim0.5\%$ in all cases.}
%\begin{ruledtabular}
\begin{tabular}{l c c c c c}
\hline\hline
Machine & $E$[GeV] & $N$[$10^{10}$] &$\langle a\rangle$  & $\langle b\rangle$
   & $\langle\zeta_x\rangle$  \\ \hline\hline
KEK's ATF & 1.4 & .9 & .01 & .10  & .15  \\
NLC & 2.0 & .75 & .01 & .20 & .40 \\
ALS & 1.0 & 5. & .015 & .25 & .15 \\ \hline\hline
\end{tabular}\vspace{-4mm}
%\end{ruledtabular}
\end{center}
\end{table}

Let us limit consideration to high energies, specifically let us assume
$a$,$b\ll1$ (if
the beam is cooler longitudinally than transversely, then this is satisfied).
We note that all 3 rings in Table~1, on average,
satisfy this condition reasonably well.
Assuming this condition,
the 2nd term in the braces of
Eq.~\ref{BM_eq} is small compared to the first term, and we drop it.
Our second assumption is to drop
off-diagonal terms (let $\zeta=0$),
and then all matrices will be diagonal.

\begin{figure}[htb]
\centering
\includegraphics*[width=70mm]{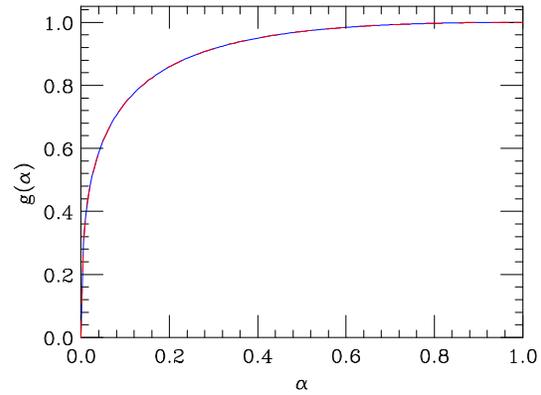}
\caption{
The auxiliary function $g(\alpha)$
(solid curve) and
the approximation, $g=\alpha^{(0.021-0.044\ln\alpha)}$ (dashes).
 \vspace{-5mm}}\label{hfun_fi}
\end{figure}

Simplifying the remaining integral by applying
the high energy assumption
we finally obtain
\begin{equation}
{1\over T_p}\approx  {r_0^2 cN({\rm log})\over
16\gamma^3\epsilon_x^{3/4}\epsilon_y^{3/4}\sigma_s\sigma_p^3}
\left<\sigma_H\,
g(a/b)\,\left({\beta_x\beta_y}\right)^{-1/4}\right>
\ , \label{BM_approx2_eq}
\end{equation}
with
\begin{equation}
g(\alpha)={2\sqrt{\alpha}\over\pi}\int_0^\infty
{du\over\sqrt{1+u^2}\sqrt{\alpha^2+u^2}}\quad.
\end{equation}
A plot of $g(\alpha)$ over the interval [$0<\alpha<1$] is given in
Fig.~\ref{hfun_fi}; to obtain the results for $\alpha>1$, note
that $g(\alpha)=g(1/\alpha)$. A fit to $g$,
\begin{equation}
g(\alpha)\approx\alpha^{(0.021-0.044\ln\alpha)}\quad\quad
[{\rm for}\ 0.01<\alpha<1]
\quad,
\end{equation}
is given by the
dashes in Fig.~\ref{hfun_fi}.
The fit has a maximum error of 1.5\% over [$0.02\leq\alpha\leq1$].

Similarly, beginning with
the 2nd and 3rd of Eqs.~\ref{BM_eq},
we obtain
\begin{equation}
{1\over T_{x,y}} \approx {\sigma_p^2\langle{\cal H}_{x,y}
\rangle\over\epsilon_{x,y}}{1\over T_p}
\quad.\label{BM_approxxy_eq}
\end{equation}
Our approximate IBS solution is Eqs.~\ref{BM_approx2_eq},\ref{BM_approxxy_eq}.
Note that
Parzen's high energy formula is a similar,
though more approximate, result to that given here\cite{Parzen:87};
and Raubenheimer's approximation
is Eq.~\ref{BM_approx2_eq}, with $g(a/b)\sigma_H/\sigma_p$ replaced by ${1\over2}$,
and Eqs.~\ref{BM_approxxy_eq} exactly as given here\cite{Raubenheimer:91}.

Note that the beam properties in Eqs.~\ref{BM_approx2_eq},\ref{BM_approxxy_eq},
need to be the self-consistent values.
Thus, for example, to find the steady-state growth rates in electron
machines, iteration will be required\cite{Piwinski:99}.
Note also that these equations
assume that the zero-current vertical emittance is due
mainly to vertical dispersion caused by orbit errors; if it is due mainly to
(weak) $x$-$y$ coupling we
let ${\cal H}_y=0$, drop the $1/T_y$ equation, and let
$\epsilon_y=\kappa\epsilon_x$, with $\kappa$ the coupling factor\cite{ATF:02B}.

\begin{figure}[htb]
\centering
\includegraphics*[width=70mm]{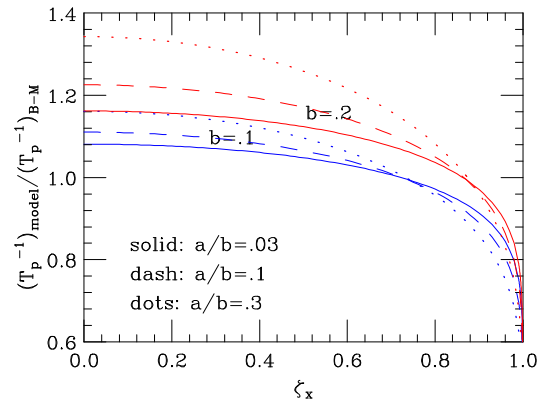}
\caption{
The ratio of {\it local} growth rates in $p$
%$(T_p^{-1})_{model}/(T_p^{-1})_{B-M}$
as function of $\zeta_x$,
for $b=0.1$ (blue) and $b=0.2$ (red) [$\zeta_y=0$].
 \vspace{-7mm}}\label{phi_fi}
\end{figure}

What sort of error does our model produce? Consider a
position in the ring where $\zeta_y=0$.
In Fig.~\ref{phi_fi} we plot the ratio of the {\it local}
growth rate $T_p^{-1}$ as given by
our model to that given by Eq.~\ref{BM_eq} as function
of $\zeta_x$, for example combinations of $a$ and $b$.
We see that
for $\zeta_x\lesssim\sqrt{b}e^{(1-\sqrt{b})}$
(which is typically true in storage rings)
%except close to $\zeta_x=1$ (a region not likely
%to be representative),
the dependance on $\zeta_x$ is weak and can be ignored.
%Beyond that
In this region we see that the model approaches B-M from above as
$a$,$b\rightarrow0$. Finally, adding small $\zeta_y\neq0$ will
reduce slightly the ratio of Fig.~\ref{phi_fi}.

%\section{COMPARISON TO THE PIWINSKI SOLUTION}
\section{COMPARISON TO PIWINSKI}

\subsection{The Standard Piwinski Solution\cite{Piwinski:99}}

The standard Piwinski solution is
\begin{eqnarray}
{1\over T_p}&=&A\left< {\sigma_h^2\over\sigma_p^2}f(\tilde a,\tilde b,q)\right>\nonumber\\
{1\over T_x}&=&A\left< f({1\over \tilde a},{\tilde b\over \tilde a},{q\over \tilde a})
+{\eta_x^2\sigma_h^2\over\beta_x\epsilon_x}f(\tilde a,\tilde b,q)\right>\nonumber\\
{1\over T_y}&=& A\left< f({1\over \tilde b},{\tilde a\over \tilde b},{q\over \tilde b})
+{\eta_y^2\sigma_h^2\over\beta_y\epsilon_y}f(\tilde a,\tilde b,q)\right>\ ;
\label{tpxy_P_eq}
\end{eqnarray}
%Parameters are:
\begin{equation}
{1\over\sigma_h^2}= {1\over\sigma_p^2} +
{\eta_x^2\over\beta_x\epsilon_x} +
{\eta_y^2\over\beta_y\epsilon_y}\quad,
\end{equation}
\begin{equation}
\tilde a={\sigma_h\over\gamma}\sqrt{\beta_x\over\epsilon_x},\quad
\tilde b={\sigma_h\over\gamma}\sqrt{\beta_y\over\epsilon_y},\quad
q=\sigma_h\beta\sqrt{{2d\over r_0}}\quad;
\end{equation}
the function $f$ is given by:
\begin{eqnarray}
f(\tilde a,\tilde b,q)&=&8\pi\int_0^1du\,{1-3u^2\over PQ}\ \times\nonumber\\
&  \times &\left\{2\ln\left[{q\over2}\left({1\over P}+
{1\over Q}\right)\right] -0.577\ldots\right\}\label{fabq_eq}
\end{eqnarray}
%where
\begin{equation}
P^2= \tilde a^2+(1-\tilde a^2)u^2,\quad\quad Q^2= \tilde b^2+(1-\tilde b^2)u^2\ .
\end{equation}
The parameter $d$ functions as a maximum impact parameter,
 and is normally taken as the vertical beam
size.\vspace{-1.5mm}

\subsection{Comparison of Modified Piwinski to the B-M Solution at
High Energies}

We note that Piwinski's result depends on $\eta^2/\beta$, and not
on $\cal H$ and $\phi$, as the B-M result does. This may suffice
for rings with $\langle{\cal H}\rangle\approx\langle\eta^2/\beta\rangle$.
For a general comparison, however,
let us consider a formulation that we call the {\it modified} Piwinski solution.
It is the standard version of Piwinski, but with $\eta^2/\beta$
replaced by ${\cal H}$ ({\it i.e.} $\tilde a$, $\tilde b$, $\sigma_h$,
become $a$, $b$, $\sigma_H$, respectively).
%First, consider
%Piwinski's formula for $1/T_p$, Eq.~\ref{tp},
%but with these substitutions.

Let us consider high energy beams, {\it i.e.} let $a$,$b\ll1$:
First, notice that in the integral of the auxiliary function
$f$ (Eq.~\ref{fabq_eq}):
the $-0.577$ can be replaced by 0; the $-3u^2$ in
the numerator can be set to 0;
$P$ ($Q$) can be replaced by $\sqrt{a^2+u^2}$ ($\sqrt{b^2+u^2}$).
The first term in the braces can be approximated by a constant and
then be pulled out of the integral; it becomes the effective Coulomb
log factor.
Note that for the proper choice of the Piwinski parameter $d$,
the effective Coulomb log can be made the same as the B-M parameter $({\rm log})$.
For flat beams ($a\ll b$), the Coulomb log of Piwinski becomes
$({\rm log})=
\ln{[d\sigma_H^2/(4r_0a^2)]}$.

We finally obtain, for the first of Eqs.~\ref{tpxy_P_eq},
\begin{equation}
{1\over T_p}\approx  {r_0^2 cN({\rm log})\over
16\gamma^3\epsilon_x^{3/4}\epsilon_y^{3/4}\sigma_s\sigma_p^3}
\left<\sigma_H\,
h(a,b)\,\left({\beta_x\beta_y}\right)^{-1/4}\right>
\ ,\label{Tpapprox_eq}
\end{equation}
with
\begin{equation}
h(a,b)={2\sqrt{ab}\over\pi}\int_0^1
{du\over\sqrt{a^2+u^2}\sqrt{b^2+u^2}}\quad.
\end{equation}
We see that the the approximate equation for $1/T_p$ for high
energy beams according to modified Piwinski is the same as that
for B-M, except that $h(a,b)$ replaces $g(a/b)$. But for
$a$,$b$ small, $h(a,b)\approx g(a/b)$, and the Piwinski result approaches
the B-M result.
For example, for the ATF with
$\epsilon_y/\epsilon_x\sim0.01$, $a\sim0.01$, $a/b\sim0.1$, and
$h(a,b)/g(a/b)=0.97$; the agreement is quite good.

Finally, for the relation between the transverse to longitudinal growth
rates according to modified Piwinski:
note that for non-zero vertical dispersion the second term
in the brackets of Eqs.~\ref{tpxy_P_eq}
(but with $\eta^2_{x,y}/\beta_{x,y}$
replaced by ${\cal H}_{x,y}$), will tend to dominate over
the first term, and the
results become the same as
for the B-M method.

In summary, we have shown that for high energy beams
($a$,$b\ll1$),
in normal rings ($\zeta$ not very close to 1):
if the parameter $d$ in P is chosen to give the
same equivalent
Coulomb log as in B-M,
then the {\it modified} Piwinski solution agrees
with the Bjorken-Mtingwa solution.\vspace{-3mm}
%Finally, if we replace $\eta^2/\beta$ by ${\cal H}$ in
%Eqs.~\ref{tx},\ref{ty}, and if $\eta_y$ is not small, we arrive at the
%second of Eqs.~\ref{Tor_eq}.
%represents a cut-off for
%the IBS force, which Piwinski says should be taken as the vertical
%beam size, but he also points out that the results are not
%very sensitive to exactly what is chosen for this parameter.

%\begin{equation}
%\sigma^2_{x\beta,y\beta,p}={\sigma^2_{x\beta0,y\beta0,p0}\over
%1-\tau_{x,y,p}/T_{x,y,p}}
%\end{equation}

\section{NUMERICAL COMPARISON\cite{ATF:02B}}

We consider a numerical comparison between results of the general
B-M method, the modified Piwinski method,
and Eqs.~\ref{BM_approx2_eq},\ref{BM_approxxy_eq}.
The example is the ATF ring with no coupling;
% and vertical dispersion
%due to random orbit errors.
to generate vertical errors, magnets were randomly offset by 15~$\mu$m, and
the closed orbit was found.
For this example
$\langle{\cal H}_y\rangle=17$~$\mu$m, yielding a zero-current emittance
ratio of 0.7\%; the beam current is 3.1~mA. The steady-state growth rates
according to the 3 methods are given in Table~2. We note that the
Piwinski results are 4.5\% low, and the results of
Eqs.~\ref{BM_approx2_eq},\ref{BM_approxxy_eq},
agree very well with those of B-M. Additionally, note that, not only the
(averaged) growth rates, but even the {\it local} growth
rates around the ring agree well
for the three cases.
Finally, note that for coupling dominated
NLC,  ALS examples ($\kappa=0.5\%$, see Table 1) the error
in the steady-state
growth rates ($T_p^{-1}$,$T_x^{-1}$) obtained with the model is
(12\%,2\%), (7\%,0\%), respectively.\vspace{-4mm}

\begin{table}[htb]
\begin{center}
\caption{
Steady-state IBS growth rates (in [s$^{-1}$])
for an ATF example with vertical dispersion
due to random errors.}
%\begin{ruledtabular}
\begin{tabular}{l c c c}
\hline\hline
Method & $1/T_p$  & $1/T_x$  & $1/T_y$ \\ \hline\hline
Modified Piwinski & 25.9 & 24.7  & 18.5 \\
Bjorken-Mtingwa & 27.0 & 26.0 & 19.4\\
Eqs.~\ref{BM_approx2_eq},\ref{BM_approxxy_eq} & 27.4 & 26.0 & 19.4\\ \hline\hline
\end{tabular}\vspace{-4mm}
%\end{ruledtabular}
\end{center}
\end{table}

%\begin{acknowledgments}
The author thanks A.~Piwinski, K. Kubo and other coauthors of
Ref.~\cite{ATF:02B}
%We thank C.~Steier for providing us with the ALS lattice to compare with.
%One author (K.B.) also
for help in understanding
IBS theory; K.~Kubo, A.~Wolski, C.~Steier, for supplying the lattices
of the ATF, NLC, ALS rings, respectively.\vspace{-2mm}

%\end{acknowledgments}


\begin{thebibliography}{99}

\bibitem{Bhat:99}
C. Bhat, {\it et al}, {\it Proc. PAC99}, New York (1999)  3155.\vspace{-1.2mm}

\bibitem{RHIC:01}
W. Fischer, {\it et al}, {\it Proc. PAC2001}, Chicago (2001)  2857.\vspace{-1.2mm}

\bibitem{ATF:02B}
K. Bane, {\it et al}, SLAC-PUB-9227, May 2002.\vspace{-1.2mm}

\bibitem{Piwinski:74}
A. Piwinski, Tech. Rep. HEAC 74, Stanford, 1974.\vspace{-1.2mm}

\bibitem{Martini:84}
M. Martini, Tech. Rep. PS/84-9(AA), CERN, 1984.\vspace{-1.2mm}

\bibitem{Piwinski:99}
A. Piwinksi, in {\it Handbook of Accelerator Physics},
World Scientific (1999) 125.\vspace{-1.2mm}

\bibitem{Bjorken:83}
J. Bjorken and S. Mtingwa, {\it Part. Accel.}, {\bf 13} (1983) 115.\vspace{-1.2mm}

\bibitem{Piwinski:p}
A. Piwinski, private communication.\vspace{-1.2mm}

\bibitem{Parzen:87}
G. Parzen, {\it Nucl. Instr. Meth.}, {\bf A256} (1987) 231.\vspace{-1.2mm}

\bibitem{LeDuff:89}
J. Le Duff, {\it Proc. of CERN Accel. School} (1989) 114.\vspace{-1.2mm}

\bibitem{Raubenheimer:91}
T. Raubenheimer, SLAC-R-387, PhD thesis, 1991, Sec. 2.3.1.\vspace{-1.2mm}

\bibitem{Wei:93}
J. Wei, {\it Proc. PAC93}, Washington, D.C. (1993)  3651.\vspace{-1.2mm}

\bibitem{Kubo:01b}
K. Kubo and K. Oide, {\it PRST-AB}, {\bf 4} (2001) 124401.\vspace{-1.2mm}



\end{thebibliography}
\end{document}